\documentclass[aps,twocolumn,showpacs]{revtex4}
\usepackage{graphicx}% Include figure files
\usepackage{dcolumn}% Align table columns on decimal point
\usepackage{bm,amsmath,amssymb}% bold math
\begin{document}

\title{Dynamical behavior of interacting dark energy in loop quantum cosmology}
\author{Kui Xiao}
\email{87xiaokui@mail.bnu.edu.cn}
  \affiliation{Department of Physics, Beijing Normal University, Beijing 100875, China}
\author{Jian-Yang Zhu}
  \email{zhujy@bnu.edu.cn}
  \affiliation{Department of Physics, Beijing Normal University, Beijing 100875, China}

\begin{abstract}
The dynamical behaviors of interacting dark
energy in loop quantum cosmology are discussed in this paper. Based on defining three
dimensionless variables, we simplify the equations of the fixed points. The fixed points
for interacting dark energy can be determined by the Friedmann equation coupled with the
dynamical equations {in Einstein cosmology}. But in loop quantum cosmology, besides
the Friedmann equation, the conversation equation also give a constrain on the fixed points.
The difference of stability properties for the fixed points in loop quantum cosmology and the
ones in Einstein cosmology also have been discussed.
\end{abstract}
\maketitle

\section{Introduction}
Dark energy \cite{Edmund-1-1} is introduced to explain the expanding
of our universe. We still have very little knowledge about it.
One of the more interested candidate is cosmological constant with
the equation of state parameter $\omega_{DE}=-1$. But the physical
interpretation of the vacuum energy density is still faultiness. And
the problem that why the dark energy and dark matter are at the same
order today is still unknown, this is so-called coincidence problem.
To overcome this problem, many dynamical dark energy models have been
introduced to replace the cosmological constant. But there are still many
other approaches to explain the coincidence problem. An interesting
proposal is that the dark energy is interacting with dark matter.
Many interacting models have been studied. The interacting term
always is the function of the Hubble parameter $H$, the density of
dark energy $\rho_{DE}$ and the density of the dark matter $\rho_{DM}$
(for example, please see the second section of \cite{Quartin-1-2} and the references of it),
or the function of $\Gamma_x$, $\rho_{DE}$ and $\rho_{DM}$
\cite{Turner-1-1,Malik-1-1,Setare-1-1,Setare-1-2,Gabriela-1-5}.

A general interaction between dark energy and dark matter is
described by
\begin{eqnarray}\label{be}
  &&\dot{\rho}_{DE}+3H\rho_{DE}(1+\omega_{DE})=Q, \\
  &&\dot{\rho}_{DM}+3H\rho_{DM}=-Q.
\end{eqnarray}
in which $\omega_{DE}$ is the equation of state parameter of dark energy. A
constant $\omega_{DE}$ and $\omega_{DE}=-1$ crossing were studied by many
authors \cite{Sadjadi-1-3, Campo-1-4,Gabriela-1-5}. In this paper
we just consider $\omega_{DE}$ is a negative constant. $Q$ is the
interacting term. And we will discuss two models:
\begin{eqnarray}
  &&\rm{Model\, I}\quad Q_1=3HC_1\rho_{DE}+3HC_2\rho_{DM},\label{model I}\\
  &&\rm{Model\, II}\quad
  Q_2=3^{2-n}CH^{3-2n}\rho_{DM}^n,\label{model II}
\end{eqnarray}
in which $C_1,C_2,C$ are some constants. Model I has been discussed by
\cite{Quartin-1-2,Gabriela-1-5} in Einstein cosmology (EC), they
considered that $C_1\neq C_2$. But more special cases also have been
studied, e.g., $C_1=C_2$ \cite{Chimento-1-6}, $C_1=0$
\cite{Chen-1-7}, $C_2=0$ \cite{Andrew-1-8}, in EC. For more simplicity, we will discuss $C_1=C_2=C$
for the first model. The second model
also has been discussed in EC
\cite{Chen-1-9,Chen-1-10}, $n$ is constant, and we will
discuss $n=2$. We will introduce three
 dimensionless variables to study those models in loop quantum cosmology (LQC),
 and discuss the difference and the common properties for dynamical system between LQC and in EC.

LQC \cite{Bojowald-1-11,Ashtekar-1-12,Singh-1-13} is a canonical
quantization of homogeneous spactimes based upon techniques used in
loop quantum gravity. Due to the homogeneous and isotropic
spacetime, the phase space of LQC is simplifier then loop quantum
gravity, e.g., the connection is determined by a single parameter
called $c$ and the triad is determined by $p$. The variables $c$ and
$p$ are canonically conjugate with Poisson bracket
$\{c,p\}=\frac{\gamma\kappa}{3}$, in which $\gamma$ is the
Barbero-Immirzi parameter. In LQC scenario, the initial singularity
is instead of a bounce. So thanks for the quantum effect, the
universe is initially contracting phase with the minimal but not
zero volume, and then the quantum effect drives it to the expanding
phase. In the effecitve scenario, the behavior of LQC also can be
well described, e.g., the big crunch instead of the big bounce
\cite{Jakub-1-14,Ding-1-15}. People always consider two kinds of
correction of effective LQC, one is inverse volume correction, the
other one is holonomy correction. In this paper, we just discuss the
holonomy correction. In this effective LQC, the Friedmann equation
is modified (see Eqs.(\ref{Fri})). It is easy to get the difference
between the Friedmann equation in EC and the one in LQC: it adds a
$-\frac{\rho^2}{\rho_c}$ term to standard Friedmann equation which
essentially encodes the discrete quantum geometric nature of
spacetime \cite{Ashtekar-1-16,Ashtekar-1-17}. This means that the
energy density will have a critical value, when the energy density
is very nearly the critical density $\rho_c$, the cosmology will
have a bounce and then oscillates forever. Due to the modification
of the Friedmann equation, the dynamical behaviors of dark energy
and dark matter are very different from the ones in EC. This factor
was showed by many authors, for example, the phantom field
\cite{Samart-1-18}, the quintom field \cite{Hao-1-19}, the
interacting phantom field \cite{Fu-1-20}, the interacting dark
energy with the interacting term
$Q=3cH\rho_{DE}^\alpha\rho_{DM}^{1-\alpha}$ \cite{Chen-1-7}, and the
dark energy interacting with cold dark matter without coupling to
the baryonic matter\cite{Li-1-21}. In this paper, we will discuss
the dynamical behavior of interacting dark energy.

This paper is organized as following. In Sec. \ref{Sec.2}, we will
review the dynamical behavior for interacting dark energy in EC. And
Sec. \ref{Sec.3} will show the dynamical behavior of interacting
dark energy in LQC. The numerical analysis will be showed in the
Sec. \ref{Sec.4}. And we get some conclusions in the last Sec.
\ref{Sec.5}. For simplicity, we set $8\pi G=1$.

\section{Dynamical behavior in EC}
\label{Sec.2}
In EC, the Friedmann equation can be written as
\begin{eqnarray}
H^2=\frac{1}{3}\rho,\label{HC}
\end{eqnarray}
in which
$\rho=\rho_{DE}+\rho_{DM}$ is the total energy density. { Differentiating Eq.(\ref{HC}) and using the conversation law of the total energy density $\dot{\rho}_t+3H(\rho+p)=0$,  we can get}
\begin{eqnarray}
\dot{H}=-\frac{1}{2}(\rho+p),\label{dHC}
\end{eqnarray}
in which $p=p_{DE}$ is
the pressure of dark energy (the pressure of dark matter is zero),
and $p_{DE}=\rho_{DE}\omega_{DE}$, with the equation of state parameter
$\omega_{DE}$. We just consider $\omega_{DE}$ is a negative
constant, then it is another parameter.
To discuss the dynamical behavior of interacting dark energy in EC, we introduce the dimensionless variables
\cite{Gabriela-1-5,Liddle-2-1}:
\begin{equation}
  x=\frac{\rho_{DE}}{3H^2},\quad y=\frac{\rho_{DM}}{3H^2},\quad
z=\frac{Q}{3H^3}.\label{xyz}
\end{equation}
Although the density of dark energy can be a negative on in the f(R)
theory \cite{Amendola-2-2}, we restrict that the
density of dark energy and dark matter both should be positive or one of then is
zero. So both of $x$ and $y$ should be non-negative.

The Friedmann equation (\ref{HC}) and Eq.(\ref{dHC}) can be rewritten as
\begin{eqnarray}
  &&x+y=1,\label{FE}\\
  &&\frac{{\dot{H}}}{H^2}=-\frac32\left(1+\frac{\omega_{DE}x}{x+y}\right)=-\frac32\left(1+{\omega_{DE}x}\right)\label{dh}.
\end{eqnarray}
The effective total equation of state parameter $\omega$ is
\begin{equation}
  \omega=\frac{\rho_{DE}\omega_{DE}}{\rho_{DE}+\rho_{DM}}=\frac{\omega_{DE} x}{x+y}=\omega_{DE} x\label{wt1}.
\end{equation}

In EC, the dynamical behavior of dimensionless variable $x,y$ are
\begin{eqnarray}
  &&x'=z-3\omega_{DE}x+3\omega_{DE}x^2,\label{x''}\\
  &&y'=-z+3\omega_{DE}xy.\label{y''}
\end{eqnarray}
in which prime denotes the derivative with respect to the e-folding
number $N=\ln a$ ($a_0=1$). Just as \cite{Gabriela-1-5} argued, if
$z=z(x,y)$, the above equations are closed and autonomous. The model
I (\ref{model I}) and II (\ref{model II}) both are in this case.

For a autonomous system $\mathbf{X}'=f(\mathbf{X})$, the fixed points $\mathbf{X}_*$ satisfy $\mathbf{X}'=0$. So, setting $x'=0,y'=0$ of Eqs.(\ref{x''},\ref{y''}) and simplifying them, it is easy to get the fixed points satisfy:
\begin{subequations}\label{ca}
 \begin{eqnarray}
  && x_*+y_*=1,\\
    &&-3\omega_{DE}x_*(1-x_*)+z_*=0,
  \end{eqnarray}
  \end{subequations}
in which $z_*$ is the value of $z$ in the fixed points. If $x_*=0$, the second subequation of Eq.(\ref{ca}) implies that $z_*=0$. This is a point for pure dark matter dominated \cite{Gabriela-1-5}, we will not consider this condition.
In this paper, we will discuss two interacting model: $Q_1=3HC_1\rho_{DE}+3HC_2\rho_{DM}$, and $Q_2=3^{2-n}H^{3-2n}C\rho_{DM}^n$.
 $C_1,$ $C_2$ and $C$ are some constants. For simplicity, we will consider $C_1=C_2=C$ and $n=2$ for those two models. Those models have been studied in EC by many authors \cite{Quartin-1-2,Sadjadi-1-3, Campo-1-4,Gabriela-1-5,Chimento-1-6,Chen-1-7,Andrew-1-8,Chen-1-9,Chen-1-10}. But for convenience to compare the dynamical behavior of interacting dark energy in LQC and the ones in EC, we review the dynamical behaviors of dark energy with different interacting
term in EC in the next two subsections.
\subsection{Model I:  $Q=3HC_1\rho_{DE}+3HC_2\rho_{DM}$}
%\label{Sec.2-A}
This interacting term has been discussed in many
papers \cite{Quartin-1-2,Sadjadi-1-3,
Campo-1-4,Gabriela-1-5,Chimento-1-6,Chen-1-7,Andrew-1-8}. In this
subsection, for more simplicity, we just consider $C_1=C_2=C$, the
dimensionless interacting term is $z=3C(x+y)$. Eq.(\ref{ca}) can be
rewritten as
\begin{eqnarray}
   -\omega_{DE}x_*(1-x_*)+C=0,\label{1ca}
   \end{eqnarray}
It is easy to get the fixed points $(x_*,y_*)$ from Eq.(\ref{1ca}):
\begin{eqnarray}
  {Point\, A}_{1}:\, && \left(\frac{1+\sqrt{1-4\frac {C}{\omega_{DE}}}}{2}, \frac{1-\sqrt{1-4\frac {C}{\omega_{DE}}}}{2}\right),\\
  {Point\, B}_{1}:\, && \left(\frac{1-\sqrt{1-4\frac {C}{\omega_{DE}}}}{2}, \frac{1+\sqrt{1-4\frac {C}{\omega_{DE}}}}{2}\right).
\end{eqnarray}

In order to determine the stability properties of the fixed points, it is necessary to expand the autonomous system $\mathbf{X}'=f(\mathbf{X})$ around the fixed points $\mathbf{X}_*$, setting $\mathbf{X}=\mathbf{X}_*+\mathbf{U}$ with $\mathbf{U}$ the perturbations of the variables considered as a column vector. Thus, for each fixed point, one can expand the equation for the perturbations up to the first order as $\mathbf{U}'=\mathbf{M}\cdot \mathbf{U}$, where the matrix $\mathbf{M}$ contains the coefficients of the perturbation equations. The stability property for each fixed point is determined by the eigenvalues of $\mathbf{M}$ \cite{Chen-1-10}. According to this property, it is easy to get the eigenvalues for the autonomous system:
 \begin{eqnarray}
  {Point\, A}_{1}:\quad && \lambda_1=\frac 32\sqrt{\omega_{DE}(\omega_{DE}-4C)}+\frac32\omega_{DE},\nonumber\\&& \lambda_2=3\sqrt{\omega_{DE}(\omega_{DE}-4C)}, \\
  {Point\, B}_{1}:\quad && \lambda_1= -\frac 32\sqrt{\omega_{DE}(\omega_{DE}-4C)}+\frac32\omega_{DE},\nonumber\\&&
  \lambda_2=-3\sqrt{\omega_{DE}(\omega_{DE}-4C)}.
\end{eqnarray}
The stability property for each fixed point is determined by the sign of $\lambda_1,\lambda_2$. The fixed point is a stable if both of $\lambda_1,\lambda_2$ are negative, and unstable if both of them are negative, if $\lambda_1,\lambda_2$ have not the same sign, then the fixed point is a saddle point.
To ensure that $0<x_*,y_*<1$, $0<4\frac{C}{\omega_{DE}}<1$ should be satisfied. For Point A$_{1}$, it is always a saddle point for $\lambda_1<0$ and $\lambda_2>0$. For Point B$_{1}$, it is a stable point always satisfied, for both of $\lambda_1, \lambda_2$ always are negative if $0<4\frac{C}{\omega_{DE}}<1$ is held. If total equation of state parameter $\omega=\frac{\omega_{DE}-\omega_{DE}\sqrt{1-4\frac C\omega_{DE}}}{2}$ is in the regions of $-1<\omega<-\frac13$, the universe will enter an accelerated stage, then Point B$_{1}$ is an accelerated scaling attractor, this has been shown in the \cite{Quartin-1-2}. If  $\omega<-1$, it is easy to get $\dot{H}>0$ from Eq.(\ref{dh}), the universe will have a big rip.

\subsection{Model II: $Q=3^{2-n}H^{3-2n}C\rho_{DM}^n$}
%\label{Sec.2-B}
This interacting term has been discussed by
\cite{Chen-1-9,Chen-1-10}. In this section, we just consider $n=2$,
the dimensionless variable is $z=3Cy^2$, then Eqs.(\ref{ca}) can be
rewritten as
\begin{eqnarray}
  -\omega_{DE}x_*(1-x_*)+C(1-x_*)^2=0.
\end{eqnarray}
The fixed points $(x_*,y_*)$ for this model are
\begin{eqnarray}
{Point\,
A}_{2}:&& \quad\left(1,0\right),\\
{Point\, B}_{2}:&& \quad\left(\frac{C}{\omega_{DE}+C},\frac{\omega_{DE}}{\omega_{DE}+C} \right).
\end{eqnarray}
And the eigenvalues are
\begin{eqnarray}
{Point\,
A}_{2}:&& \quad\lambda_1=3\omega_{DE},\quad \lambda_2=3\omega_{DE},\\
{Point\, B}_{2}:&& \quad \lambda_1=-3\omega_{DE}, \quad \lambda_2=\frac{3\omega_{DE}C}{C+\omega_{DE}}.
\end{eqnarray}
For Point A$_{2}$, the condition of $x_*,y_*>0$ and $\omega_{DE}<0$ constrain that $C<0$ and $\omega_{DE}+C<0$ should be held. It is always a stable point for $\omega_{DE}<0$. The universe will entre a dark energy dominate stage, and $\omega=\omega_{DE}$. For Point B$_{2}$, $C<0$ should be satisfied, it is always a saddle point. So, for this interacting model, it has not any attractive behavior except the universe enter a dark energy dominated stage.

\section{Dynamical behavior in LQC}
\label{Sec.3}
In this section, we discuss the dynamical behavior of
interacting dark energy in LQC. As mentioned in the introduction,
{the Friedmann equation is modified in LQC and can be written as}
\begin{eqnarray}
  &&H^2=\frac{1}{3}\rho\left(1-\frac{\rho}{\rho_c} \right),\label{Fri}
\end{eqnarray}
in which $\rho_c=\frac{3}{\kappa\gamma^2\alpha\ell_{\rm pl}^2}$. {Differentiating Eq.(\ref{Fri}) and using the conversation law of the total energy density $\dot{\rho}_t+3H(\rho+p)=0$,  one can get}
\begin{eqnarray}
  &&\dot{H}=-\frac12(\rho+p)\left(1-\frac{2\rho}{\rho_c}
  \right)\label{dH},
\end{eqnarray}

Considering the dimensionless variables (\ref{xyz}), {we can rewritten Eqs.(\ref{Fri},\ref{dH}) as}
\begin{eqnarray}
    && (x+y)\left[1-\frac{3H^2}{\rho_c}(x+y)\right]=1,\label{Fri-xy}\\
  &&  \frac{\dot{H}}{H^2}=-\frac{3}{2}\left[1+\frac{x\omega_{DE}}{x+y}
  \right](2-x-y).\label{Raych-xy}
\end{eqnarray}

The total effective equation of state parameter $\omega$ is
\begin{equation}
  \omega=\frac{\rho_{DE}\omega_{DE}}{\rho_{DE}+\rho_{DM}}=\frac{\omega_{DE} x}{x+y}.\label{wt}
\end{equation}

Considering the dimensionless variables, the dynamical equations for
our interesting system can be obtained:
\begin{eqnarray}
 &&  x'=z-3x(1+\omega_{DE})\nonumber\\
 &&\quad +3x\left(1+\frac{x\omega_{DE}}{x+y}
  \right)(2-x-y),\label{x'}\\
&&   y'=-z-3y+3y\left(1+\frac{x\omega_{DE}}{x+y}
  \right)(2-x-y),\label{y'}
\end{eqnarray}
in which prime denotes the derivative with respect to the e-folding
number $N=\ln a$ ($a_0=1$).

The fixed points $(x_*,y_*)$ satisfy $x'=0,y'=0$. Simplifying Eqs.(\ref{x'},\ref{y'}), it is easy to get
\begin{subequations}\label{case I}
 \begin{eqnarray}
  {Case\, I} \quad && x_*+y_*=1,\\
    &&-3\omega_{DE}x_*(1-x_*)+z_*=0.
  \end{eqnarray}
  \end{subequations}
or
\begin{subequations}\label{case II}
 \begin{eqnarray}
 {Case\, II}\quad &&  x_*(1+\omega_{DE})+y_*=0,\\
   && -3x_*(1+\omega_{DE})+z_*=0.
  \end{eqnarray}
\end{subequations}
Comparing the above equations with Eqs.(\ref{ca}), we find
that the expression for the first case is as same as in the EC.
The Case II is a special case which just exists in LQC. This is because of the Friedmann equation has been modified, then the dynamical equations in LQC are different from the ones in EC (see
Eqs.(\ref{x'},\ref{y'}) and Eqs.(\ref{x''},\ref{y''})). We will discuss these two cases for fixed point in the last section.

As in EC, the values of fixed points depend
on the special form of dimensionless interacting term $z$. We will consider two models: $Q=3HC_1\rho_{DE}+3HC_1\rho_{DM}$, and
$Q=3^{2-n}H^{3-2n}C\rho_{DM}^n$, with some constants $C_1,C_2,C,n$.

\subsection{Model I:  $Q=3HC_1\rho_{DE}+3HC_2\rho_{DM}$}
%\label{Sec.3-A}
In this model, the dimensionless variable $z$ can be
written as
\begin{eqnarray}
  z=\frac{Q}{3H^3}=3C_1x+3C_2y,
\end{eqnarray}
in which $C_1,C_2$ are some constants.

Considering the above equation, Case I and Case II can be rewritten
as
\begin{eqnarray}
&&\omega_{DE}x_*^2-\omega_{DE}x_*+C_1x_*+C_2(1-x_*)=0,\label{11ca}\\
&&C_1x_*+C_2[(-x_*(1+\omega_{DE})]-x_*(1+\omega_{DE})=0.\label{11cb}
\end{eqnarray}
The first equation corresponds to Case I, and the second one
corresponds to Case II. It is easy to find that $x_*=0$ in the Case II. This means that $y_*=0$. The universe will enter other stuff dominated stage, this is beyond our interesting, so we will ignore this case in this subsection.
As the last section, we just consider $C_1=C_2=C$.

Now, we consider Case I. Eq.(\ref{11ca}) becomes
\begin{equation}
x_*^2-x_*+\frac C{\omega_{DE}}=0.
\end{equation}
Then the fixed points $(x_*,y_*)$ are
\begin{eqnarray}
  {Point\quad A}_{3}:\quad \left(\frac{1+\sqrt{1-4\frac C{\omega_{DE}}}}{2}, \frac{1-\sqrt{1-4\frac C{\omega_{DE}}}}{2}\right),\\
  {Point\quad B}_{3}:\quad \left(\frac{1-\sqrt{1-4\frac C{\omega_{DE}}}}{2}, \frac{1+\sqrt{1-4\frac
  C{\omega_{DE}}}}{2}\right).
\end{eqnarray}
The eigenvalues are
\begin{eqnarray}
     {Point\quad A}_{3}:&&\,\lambda_1=-3-\frac 32 \omega_{\rm {DE}}-\frac32\sqrt{\omega_{\rm {DE}}(\omega_{\rm {DE}}-4C)},\nonumber\\ &&\lambda_2=3\sqrt{\omega_{\rm {DE}}(\omega_{\rm {DE}}-4C)},\\
  {Point\quad B}_{3}:&&\, \lambda_1=-3-\frac 32 \omega_{\rm {DE}}+\frac32\sqrt{\omega_{\rm {DE}}(\omega_{\rm {DE}}-4C)},\nonumber\\
   &&\lambda_2=-3\sqrt{\omega_{\rm {DE}}(\omega_{\rm {DE}}-4C)}.
\end{eqnarray}
The density of dark energy and dark matter both are positive ones or
one of them is zero, this condition given a constraint on the relationship
between $\omega_{DE}$ and $C$. In this case, $\omega_{DE}<0$ and
$0<\frac{C}{\omega_{DE}}\leq \frac14$ both should be satisfied.  For Point A$_{3}$, it is a unstable point if $1+\omega_{\rm {DE}}+\omega_{\rm {DE}}C>0$ and $\omega_{\rm {DE}}<-2$, and it is a saddle point if $\omega_{\rm {DE}}>-2$ or $1+\omega_{\rm {DE}}+\omega_{\rm {DE}}C<0$ and $\omega_{\rm {DE}}<-2$. For Point B$_{3}$, it is stable point if $\omega_{DE}>-2$ and $1+\omega_{DE}+\omega_{DE}C>0$, and a saddle point if  $\omega_{DE}>-2$ and $1+\omega_{DE}+\omega_{DE}C<0$ or $\omega_{DE}<-2$.

Comparing the above fixed points and eigenvalues with the ones of
classical scenario, we find the
expressions for fixed points are same for it is not necessary to consider Case II, but the eigenvalues are
different. The reason is that the eigenvalues are determined by the dynamical
equations, the dynamical equations in LQC is different from the ones in EC. In EC, the singularity maybe happens. But in LQC, for Friedmann equation has been modified, the singularity is replaced by the bounce always. We will show this evidence in the next section.

\subsection{Model II: $Q=3^{2-n}CH^{3-2n}\rho_{DM}^n$}
%\label{Sec.3-B}
In the model, the dimensionless interacting term is
$z=3Cy^n$, then the Eqs.(\ref{case I},\ref{case II}) can be
rewritten as
\begin{eqnarray}
   &&\omega_{DE}x_*^2-\omega_{DE} x_*+C(1-x_*)^n=0\label{m-2-1},\\
  && (-1)^nC(1+\omega_{DE})^nx_*^n-x_*(1+\omega_{DE})=0\label{m-2-2}.
\end{eqnarray}
To get the values of $x_*$, we need the special value of $n$. As an example to use Case II to get the fixed points, we will consider $n=2$. Then, the dimensionless variable $z$ is $z=3Cy^2$.
In this case, Eqs.(\ref{m-2-1},\ref{m-2-2}) can be rewritten as
\begin{subequations}\label{n=2}
  \begin{gather}
\omega_{DE}x_*^2-\omega_{DE}x_*+C(1-x_*)^2=0\label{n1},\\
  Cx_*^2(1+\omega_{DE})^2-x_*(1+\omega_{DE})=0.\label{nn2}
  \end{gather}
\end{subequations}
Solving the above equations, it is easy to get the fixed points
\begin{eqnarray}
{Point\,
A}_{4}:&& \quad\left(1,0\right),\\
{Point\, B}_{4}:&& \quad\left(\frac{C}{\omega_{DE}+C},\frac{\omega_{DE}}{\omega_{DE}+C} \right),\\
 {Point\, C}_{4}:&& \quad\left(\frac{1}{C(1+\omega_{DE})},-\frac{1}{C}\right).
\end{eqnarray}
Noticed that $x_*=0$ is also the solution of Eq.(\ref{nn2}), but considering the
first equation of Eq.(\ref{case II}), $y_*$ should be zero when $x_*=0$, the
universe will enter other stuff dominated, this is beyond our
interesting.

The eigenvalues for above points are
\begin{eqnarray}
{Point\,
A}_{4}:&& \,\lambda_1=-3(1+\omega_{DE}),\quad \lambda_2=3\omega_{DE},\\
{Point\, B}_{4}:
&& \lambda_1=-3\omega_{DE},\nonumber\\ &&\lambda_2=-3\frac{C+C\omega_{DE}+\omega_{DE}}{\omega_{DE}+C}, \\
{Point\, C}_{4}:&& \, \lambda_1=-\frac 32
 \omega_{DE}\nonumber\\ &&-\frac32\frac{\sqrt{\omega_{DE}^2C^2-4C^2-4\omega_{DE}C^2-4C\omega_{DE}}}{C},\nonumber\\
 && \lambda_2=-\frac 32 \omega_{DE}\nonumber\\ &&+\frac32\frac{\sqrt{\omega_{DE}^2C^2-4C^2-4\omega_{DE}C^2-4C\omega_{DE}}}{C}.\nonumber\\
 &&
\end{eqnarray}
For Point $\rm{A}_{4}$, the universe will enter a dark energy
dominated stage. If $-1<\omega_{DE}<-\frac13$, e.g., the dark energy is
quintessence-like type, this point is a stable point, and it is an
accelerated scaling attractor. If $\omega_{DE}<-1$, it is a saddle
point. For Point $\rm{B}_{4}$, to ensure $0<x_*,y_*<1$, $C<0$
should be satisfied. So it is a saddle point if
$C+C\omega_{DE}+\omega_{DE}<0$ and a unstable point if
$C+C\omega_{DE}+\omega_{DE}>0$. For Point $\rm{C}_{4}$, $C<0, \omega_{DE}<-1$ and
$\omega_{DE}^2C-4C-4\omega_{DE}C-4\omega_{DE}<0$ need to be satisfied, it
is a unstable point if $C+C\omega_{DE}+\omega_{DE}<0$ and a saddle point
if $C+C\omega_{DE}+\omega_{DE}>0$.

So, this model in LQC will have an attractor behavior just when the universe enter a dark energy with $-1<\omega_{DE}<-\frac13$ dominated stage. This is different from the one of EC, in which the dynamical system will have an attractor behavior when the universe is all types of dark energy dominated.
As the first model, due to the quantum modification of Friedmann
equation, the dynamcial behavior for LQC is different from the one for EC.

\section{Numerical analysis}
\label{Sec.4}
In this section, we will analysis the dynamical
behavior of interacting dark energy by using numerical tool.

\begin{figure}[h!]
\includegraphics[width=0.37\textwidth]{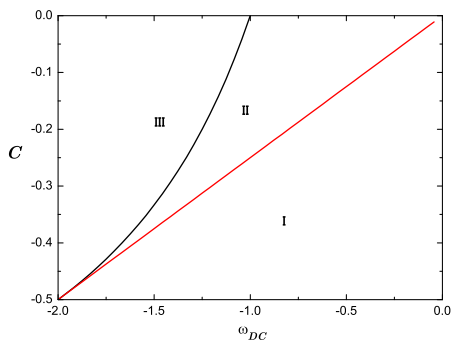}
\caption{The stable regions in the $(\omega_{DE},C)$ parameter space
for Model I. In EC, the stable regions are region
$\rm{II}+\rm{III}$. In LQC, the stable region is region II. }
\label{Fig.1}
\end{figure}

In this paper, we restricted that $0\leq x_*,y_*\leq 1$. This
restriction gives a constraint on $C,\omega_{DE}$. As discussion
before, although the expressions for the fixed points in LQC are as
same as the ones in EC for Model I, but due to the different
dynamical equations, the stability behaviors for each fixed points
in LQC are different from the ones in EC. We show the stable regions
in the parameter space $(\omega_{DE}, C)$ for the Model I in
Fig.\ref{Fig.1}. It is easy to find that the stable region for LQC
is smaller then the one for EC.

\begin{figure}[h!]
\includegraphics[width=0.37\textwidth]{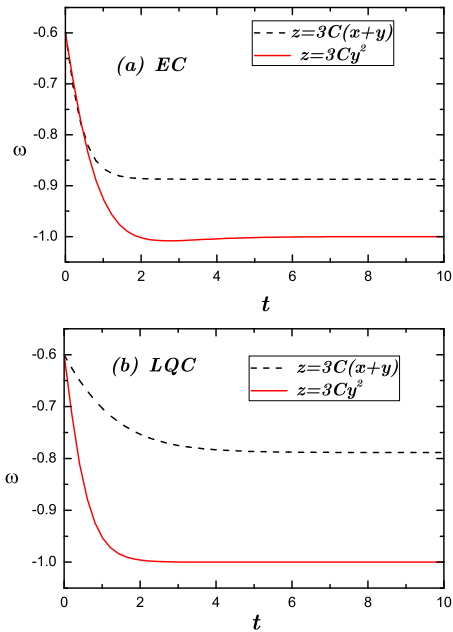}
\caption{(Color online). The evolution $\omega$ in LQC. The values
of constants are $C=-0.1,\omega_{DE}=-0.6,\rho_c=0.82$. The upper
panel is for EC, and the under one is for LQC.} \label{Fig.2}
\end{figure}

\begin{figure}[h!]
\includegraphics[width=0.37\textwidth]{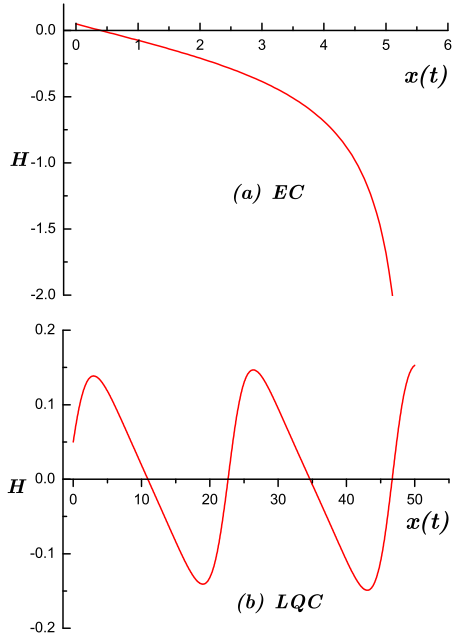}
\caption{The evolution of Hubble parameter in the Model II. The
values of constants are $C=-0.05,\omega_{DE}=-0.6,\rho_c=0.82$.The
upper panel is for EC, and the under one is for LQC.} \label{Fig.3}
\end{figure}

In Fig.\ref{Fig.2}, we plot the evolution of the total equation of
state parameter $\omega_{\rm{t}}$ (described by
Eqs.(\ref{wt1},\ref{wt})) for different models. As
Eqs.(\ref{wt1},\ref{wt}) showed, the value of
$\omega=\frac{\rho_{DE}\omega_{DE}}{\rho_{DE}+\rho_{DM}}$ is
determined by the values of $\rho_{DE}$, $\rho_{DM}$, and also
depends on the interacting term. But just as the diagram showed, the
final value of $\omega$ will tend to constant. When
$\omega_{DE}=-0.6, C=-0.1$, we can get the total effective equation
of state parameter $-1<\omega<-\frac13$ in the final data for Model
I, not only in EC, but also in LQC. But for model II, $\omega$ will
be less then $-1$, and the universe will be phantom-like dark energy
dominated. In EC, the universe will have a singularity, but in LQC,
the universe will enter an oscillatory region and will have not any
singularity, just as Fig.\ref{Fig.3} shows.

\begin{figure}[h!]
\includegraphics[width=0.37\textwidth]{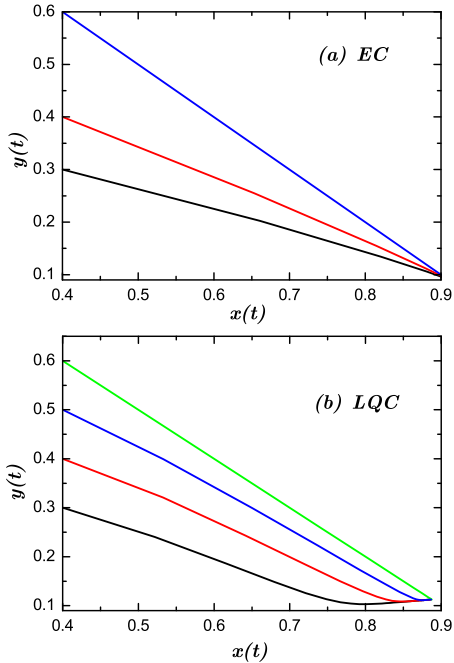}
\caption{(Color online). The phase diagrams of interacting dark
energy. The upper panel is for EC, with $\omega_{DE}=-1.2,C=-0.1$.
And the under one is for LQC, with
$\omega_{DE}=-1,C=-0.1,\rho_c=0.82$.}\label{Fig.4}
\end{figure}

Figure \ref{Fig.4} shows the stable point in the classical cosmology
and in LQC for Model I. Although the expressions of the fixed points
in the classical cosmology are as same as the ones in LQC, but due
to the dynamical equations are different, so the stable region in
$(\omega_{DE}, C)$ space is very different, just as Fig.(1) shows,
then the value of stable points for those two cosmology are
different.

\begin{figure}[h!]
\includegraphics[width=0.37\textwidth]{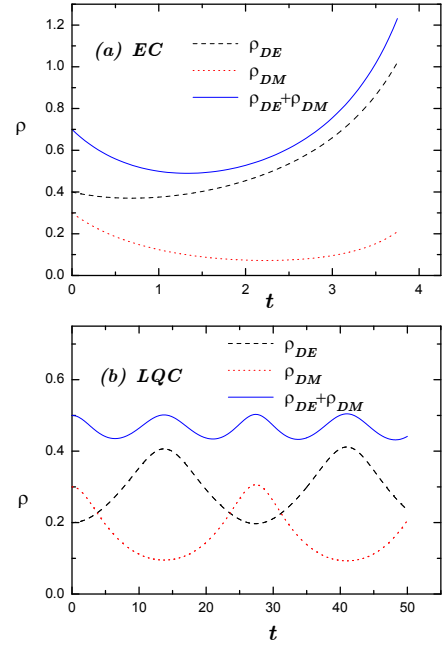}
\caption{(Color online). The evolution of energy density $\rho$. The
upper one is the evolutions of  $\rho_{DE},\rho_{DM}$ and the total
energy density  $\rho=\rho_{DE}+\rho_{DM}$ in EC. The under one is
for those variables in LQC. Both of those two diagrams are for
$\omega_{DE}=-1.6,C=-0.1$.} \label{Fig.5}
\end{figure}
In Fig.\ref{Fig.5}, we show the evolution of the densities in EC
scenario and LQC scenario. In EC, the densities will become bigger
and bigger. But in LQC scenario, the densities of dark energy and
dark matter and the total matter will not divergence. The universe
will entre a oscillating regime, and the singularity is avoided.

\section{Discussion and conclusion}
\label{Sec.5}
In this paper, based on defining some dimensionless
variables, we simplified the equations of fixed points. We found
$x_*+y_*=1$ for the fixed points should be satisfy both in EC and
LQC. It is easy to understand this condition hold in EC, for $x+y=1$
is nothing but the Friedmann equation, just as Eq.(\ref{FE}) shows.
But in LQC, if the quantum effect is dominated, it is impossible to
be satisfied for $\rho<\rho_c$. But notice that the quantum effect
should be considered just when $\rho$ is comparable to $\rho_c$. So
$x_*+y_*=1$ means that our universe is described by general
relativity very well when the dynamical system arrives its fixed
point. The energy density is very far from the one should consider
the quantum effect. It is easy to check that all fixed points of the
interacting models discussed in \cite{Chen-1-7} are determined by
the condition of Case I. Case II is just held in LQC, but we can get
this case from the conversation equation for energy density
\begin{equation}
  \dot{\rho}=-3H[\rho_{DE}(1+\omega_{DE})+\rho_{DM}].\label{bt}
\end{equation}
When $\dot{\rho}_{DE}=\dot{\rho}_{DM}=0$, this implies $\rho_{DE}(1+\omega_{DE})+\rho_{DM}=0$ and $x'=y'=0$. Although the conversation for energy density should be satisfied both in EC and LQC, $x_*(1+\omega_{DE})+y_*=0$ means that $x_*\omega_{DE}=-1$ in classical universe, but considering Eqs.(\ref{x''},\ref{y''}) and setting $x'=y'=0$, this is nothing but $x_*+y_*=1$. So, Case II is just a special condition of Case I in the region where the universe is described by general relativity. When the universe is quantum effect dominated, $x_*+y_*=1$ will not be satisfied, so, Case II is a special condition in LQC. Considering Eq.(\ref{case II}), we can find that the fixed points are described by this case is happening when $z_*\propto x_*^n$ with $n\neq 0,1$.
Notice that, $\omega_{DE}<-1$ should be satisfied in this case for $\rho_{DE},\rho_{DM}>0$. Those fixed points which are determined by Case II are staying in the quantum regions, and if they are stable points, the universe will not entre the Einstein cosmology.

Due to the Friedmann equation is modified in LQC, the dynamical equations in LQC are
different from the ones in EC. Under the help of the simplified equations of fixed points, we
discussed the dynamical behavior for two different interacting
models in LQC. We find that the stability of system dependent
on the parameters $\omega_{DE}, C$. Also, $0\leq x_*,y_*\leq 1$
gives a constraint on $\omega_{DE},C$. The attractor solutions for autonomous system in LQC are very different from the ones in EC, this fact is obvious when we compare the stable regions in $(\omega_{DE}, C)$ spaces of two conditions, although the attractor solution is lying in the stage which is described by general relativity both for LQC and EC. The values of those parameters also should
be constrained by the observation
data, it is discuss by many authors \cite{Olivares-2-2,Wang-2-3,Wang-2-4,Rowland-5-1} in EC, but it is still need more studying in LQC. Also, just as we expecting, the
unstable region in LQC, the universe will entre
an oscillatory regime and avoid the singularity which always happens
in EC.

\acknowledgments The work was supported by the National Natural
Science of China (No.10875012) and the Scientific Research
Foundation of Beijing Normal University.

\end{document}